\documentclass[final]{ws-ijmpa}

\usepackage[super,compress]{cite}
\usepackage{amsmath,amssymb,amsfonts}
\usepackage{bm,bbm}
\usepackage{graphicx}
\usepackage{esint}
\usepackage{color}
\usepackage[breaklinks]{hyperref}
\usepackage[usenames,dvipsnames]{xcolor}

\let\normalcolor\relax  

\def\cob{\color{blue}}
\def\cop{\color{Blue}}
\newcommand{\be}{\begin{equation}}
\newcommand{\ee}{\end{equation}}
\newcommand{\ba}{\begin{eqnarray}}
\newcommand{\ea}{\end{eqnarray}}
\def\bs{\begin{subequations}}
\def\es{\end{subequations}}
\def\a{\alpha}
\def\b{\beta}
\def\de{\delta}

\def\la{\lambda}

\def\e{\epsilon}

\def\vp{\varphi}

\def\cD{\mathcal{D}}

\def\cL{\mathcal{L}}

\def\cS{\mathcal{S}}

\def\bE{\mathbbm{e}}

\def\dh{d_{\rm H}}

\def\p{\partial}

\newcommand{\Eq}[1]{(\ref{#1})}

\def\cob{\color{blue}}

\newcommand{\books}[4]{\emph{#1} (#2, #3, #4)}
\newcommand{\oarX}[1]{\href{http://arxiv.org/abs/#1}{{\ttfamily\cop arXiv:#1}}}
\newcommand{\arX}[1]{\href{http://arxiv.org/abs/#1}{{\ttfamily\cop arXiv:#1}}}
\newcommand{\doin}[6]{\href{http://dx.doi.org/#1}{{\cob {\it #2 #3} {\bf #4}, #5 (#6)}}}
\newcommand{\doinn}[5]{\href{http://dx.doi.org/#1}{{\cob {\it #2} {\bf #3}, #4 (#5)}}}
\newcommand{\doij}[5]{\href{http://dx.doi.org/#1}{{\cob {\it #2} {\bf #3}, #4 (#5)}}}

\newcommand{\tia}[1]{}

\def\rme{e}
\def\rmd{d}
\def\rmi{i}

\begin{document}


\title{QUANTUM FIELD THEORY WITH VARYING COUPLINGS}
\author{GIANLUCA CALCAGNI}
\address{Instituto de Estructura de la Materia, IEM-CSIC,\\ Serrano 121, 28006 Madrid, Spain\\ calcagni@iem.cfmac.csic.es}
\author{GIUSEPPE NARDELLI}
\address{Dipartimento di Matematica e Fisica, Universit\`a Cattolica,\\ via Musei 41, 25121 Brescia, Italy}
\address{INFN Gruppo Collegato di Trento, Universit\`a di Trento,\\ 38100 Povo (Trento), Italy\\ nardelli@dmf.unicatt.it}

\maketitle


\begin{abstract}
A quantum scalar field theory with spacetime-dependent coupling is studied. Surprisingly, while translation invariance is explicitly broken in the classical theory, momentum conservation is recovered at the quantum level for some specific choice of the coupling's profile for any finite-order perturbative expansion. For one of these cases, some tree and one-loop diagrams are calculated. This is an example of a theory where violation of Lorentz symmetry is not enhanced at the quantum level. We draw some consequences for the renormalization properties of certain classes of fractional field theories.
\end{abstract}

\date{June 4, 2013}

\ccode{PACS numbers: 11.10.--z, 11.30.Cp, 11.10.Kk}



\

\centerline{\doin{10.1142/S0217751X14500122}{Int.\ J.\ Mod.\ Phys.}{A}{29}{1450012}{2014} [\arX{1306.0629}]}


\section{Introduction}

Conservation laws associated with symmetries are of fundamental importance in quantum field theory. Momentum is not conserved in systems where translation invariance is broken. In classical mechanics or classical field theory, this is somewhat inescapable due to the tight relation between symmetries and conservation laws, via Noether's theorem.

At the quantum level, the problem is much less obvious than expected. There may exist nonautonomous models avoiding the above conclusion. Namely, a classical system with a manifestly broken symmetry (translation) leads to a quantum theory where momentum conservation is perturbatively recovered, at least at finite order. Models with nonstandard momentum conservation laws often appear in quantum gravity and noncommutative spacetimes, but here we will concentrate on field-theory scenarios where the interaction effective couplings explicitly depends on spacetime coordinates. In this paper, we will make a few remarks on such class of systems, focusing in particular on the properties of some one-loop diagrams in a scalar field theory. We will find that momentum conservation is guaranteed only at a finite-order perturbative expansion. The present analysis is only preliminary, since it is not yet clear whether full resummation of the perturbative series maintains momentum conservation. Still, it may be of interest for possible applications in effective field theories with varying couplings.

The latter have been considered at a purely classical level. A popular example is electromagnetism with varying electric charge \cite{Bek82,Bek02}, later followed by Chern--Simons electrodynamics with an external vector \cite{CFJ}, gravity with a Pontryagin term \cite{LWK,JaPi3}, the electroweak sector of the Standard Model with varying gauge couplings \cite{KiM,ShB1}, chromodynamics \cite{CLV}, and grand-unification scenarios \cite{CaF,LSS,DNRV}. In almost all these cases, however, the couplings are Lorentz scalar fields endowed with kinetic terms and, therefore, such that dynamics itself determines their profiles. Here we are interested in a rather different setting, where the couplings are not dilaton-like fields but given generalized functions of the coordinates. This is done both for the sake of stating the problem as simply as possible and because of a major application of the present results, namely, to field theories on multiscale spacetimes. These have been introduced for the purpose of describing effective regimes of spacetime geometry with anomalous properties (correlation functions, spectral dimension and so on), which appear in various approaches to quantum gravity and noncommutative spacetimes \cite{ACOS,fra7} (see also Ref.\ \citen{AIP} and references therein). A somewhat similar situation occurs also in holographic descriptions of cosmological string backgrounds, where the dilaton profile is fixed \emph{a priori}; this produces non-Poincar\'e invariant effective field actions with nonautonomous potentials \cite{CSV,RoS,CRS,MRS}.

Apart from momentum conservation, there is another consequence we wish to highlight. Lorentz-symmetry violation of the present model is \emph{not} enhanced by loop effects because the effective dispersion relation is modified by terms which decay in the ultraviolet. This result is independent of the ultimate renormalization properties of the model and is in contrast with other scenarios with modified dispersion relations breaking Lorentz symmetry, where the extra terms lead to large fine tunings \cite{CPSUV,CPS}. 


\section{The Model}

The prototype of nonautonomous scenarios we shall consider is the scalar-field action
\be\label{Spsi}
S =\int_{-\infty}^{+\infty}\rmd^Dx\,\cL\,,\qquad \cL =\frac12\vp\p_\mu\p^\mu\vp-\frac12 m^2\vp^2-\frac{\la(x)}{n!}\vp^n\,,
\ee
where $D$ is the number of topological dimensions (indexed by $\mu=0,1,\dots,D-1$), $\vp$ is a Lorentz scalar, $\p^\mu=\eta^{\mu\nu}\p_\nu=\eta^{\mu\nu}\p/\p x^\nu$, $\eta={\rm diag}(-,+,\cdots,+)$ is the Minkowski metric, $m$ is the mass of the field, and $\la(x)$ is the spacetime-dependent coupling of a monomial potential ($n\geq 3$). Because of the explicit spacetime dependence, the variation of $S$ under translations is not zero and the energy-momentum tensor $T^\mu {}_{\nu} :=\de^\mu_\nu\, \cL+\p^\mu\vp\p_\nu\vp$ is not conserved \cite{frc6}: $\p_\mu T^\mu{}_\nu = \p_\nu \la\,\vp^n/n!$.

The above action is the formal rewriting of a field theory living in an anomalous spacetime with measure $\rmd^Dx\,v(x)$, where $v(x)$ is a weight encoding geometric information of the background. Such spacetimes possess a hierarchy of scales such that their Hausdorff and spectral dimensions change with the probed scale and they exhibit multiscale (in particular, multifractal) properties \cite{AIP,fra1,fra4,frc1,frc2,fra6}. In example \Eq{Spsi}, the action $S$ can be recast in terms of the fundamental field density $\phi(x)=\vp(x)/\sqrt{v(x)}$, $S=\int\rmd^Dx\, v(x)[(1/2)\phi\cD^2\phi-(1/2)m^2\phi^2-(\la_0/n!)\phi^n]$, where the kinetic operator is $\cD^2=v^{-1/2}\p_\mu\p^\mu(v^{1/2}\,\cdot\,)$, $\la_0$ is a constant and the measure $v$ determines the coupling $\la$ in \Eq{Spsi} as $\la(x)=\la_0[v(x)]^{1-n/2}$ \cite{frc6}.

The focus of this paper will be not so much on multiscale spacetimes but, rather, on the quantum theory stemming from Eq.\ \Eq{Spsi}. We shall consider the following form of the coupling:
\be\label{la}
\la(x)=\la_{\b-1}(x):=\la_0 \prod_{\mu=0}^{D-1}\left[c_\b\left|\frac{x^\mu}{\ell}\right|^{\b-1}\right]\,,
\ee
where $\la_0$, $\b$ and $c_\b$ are constants and $\ell$ is a length or time scale. In fractional models \cite{frc1}, it is common to choose $c_\b=1/\Gamma(\b)$. In this interpretation, the local coupling \Eq{la}  would correspond to an isotropic fractional measure weight $v(x)\propto \prod_\mu |x^\mu|^{\a-1}$ and a Hausdorff dimension $\dh=D\a$, where ($n\geq 3$)
\be\label{ab}
\a=\frac{n-2\b}{n-2}\,.
\ee
We will not consider this relation until the very end, when we will discuss the results. 

If the coupling is required to be locally integrable, then it follows that $\b >0$. However, in a quantum-field-theory context, it is natural to consider  $\la$ as a distribution. As such, it can be analytically continued on the whole complex $\beta$ plane except at the points $\b=-2l$, $l=0,1,2,\dots$, where the distribution  $|x|^{\b-1}$ is singular. More precisely, for each direction $x^\mu$ (denoted by $x$ for brevity) and for any  rapidly decreasing test function $\psi\in \cS(\mathbb{R})$, the distribution $|x|^{\b-1}$ is defined as $(|x|^{\b-1},\psi)=\int_0^{+\infty}\rmd x\,x^{\b-1}[\psi(x)-\psi(-x)]$ if $\b>0$. This formula can be analytically continued to  any strip $-2N-2<{\rm Re} \b <-2N$,  $N\in\mathbb{N}$, as \cite{GS}
\ba
(|x|^{\b-1},\psi)&=& \int_0^{+\infty}\rmd x\,x^{\b-1}\left[\psi(x)-\psi(-x)-2\sum_{l=0}^N\frac{\psi^{(2l)}(0)}{(2l)!}x^{2l}\right]\,,\nonumber\\
&& \qquad-2N-2<{\rm Re} \b<-2N\,,\label{strips}
\ea
where the superscript $(2l)$ indicates a derivative with respect to $x$ of order $2l$. In the complex $\b$ plane, the only singular points are at $\b=0,-2,-4,\dots$, where the distribution $|x|^{\b-1}$ has simple poles with residue \cite{GS}
\be\label{usefu}
{\rm Res}\left[|x|^{\b-1}\right]_{\b=-2l}=\frac{2}{(2l)!}\,\de^{(2l)}(x)\,,
\ee
as it can be checked by a direct inspection. The generalization to $D$ dimensions is straightforward. Incidentally, the coefficient $c_\b$  in the fractional measure is  $1/\Gamma(\b)$, which vanishes at the singular points $\b= -2 l$,  making  the distribution $\lambda_{\b -1}$ an  entire function  over the whole complex $\b$ plane. It should be stressed, however, that in what follows little will rely on any specific choice of $c_\b$.
 

\section{Vertex and Momentum Conservation}

At the classical level, momentum is not conserved, since $\p_\mu T^\mu{}_\nu\neq 0$ \cite{frc6}. Therefore, one may wonder whether Feynman rules can be defined at all. Translation noninvariant terms involve only the interaction, so that the free scalar theory is standard. The free partition function is
\be
 Z_0[J]=\exp\left[\frac{\rmi}{2}\int\rmd^Dx\,\rmd^Dy\, J(x) G_0(x-y) J(y)\right],
 \ee
where 
\be
G_0(x-y) =\int \frac{\rmd^Dk}{(2\pi)^D}\,\frac{\rme^{\rmi k\cdot(y-x)}}{k^2+ m^2 -\rmi \e}=:\int \frac{\rmd^Dk}{(2\pi)^D}\,\rme^{\rmi k\cdot(y-x)} \tilde G_0(k^2)
\ee
is the free Feynman propagator. The complete partition function can be written as a functional operator acting on the free partition function, 
\be
\label{zj}
Z[J]=\exp\left\{\frac{\rmi}{n!}\int\rmd^Dx\,\la_{\b-1}(x)\left[\frac{1}{\rmi}\frac{\de}{\de J(x)}\right]^n\right\}Z_0[J]\,.
\ee
 As in ordinary field theory, in this expression the coupling function $\la_{\b-1}$ is the bare one. Due to the fact that the shape of the vertices will not be reproduced order by order, in general the relation between bare and dressed coupling will be nonlinear. This complicates the renormalization procedure that, if it exists, it will follow nonstandard schemes. This is a problem we will not analyze here. 

Another important remark regards the perturbative expansion. The latter is thought with respect to the constant $\la_0\ll 1$, but in practice it is indistinguishable from an expansion in $\la_{\b-1}(x)$. However, when regarded as an expansion in $\la_{\b-1}$ one may be worried that a perturbative truncation is unjustified in regions of space where the coupling $\la_{\b-1}(x)$ is large; in the present case, this happens when $x\to\infty$. Yet, the spacetime dependence of the coupling is nothing but a nontrivial measure appended to the integral in \Eq{zj} and the limit $x\to\infty$ in $\la_{\b-1}(x)$ must be interpreted in the sense of distributions, not of ordinary functions \cite{frc6}. This implies that, however nonstandard the functional form of the partition function \Eq{zj} may be, one can always make sense of a perturbative expansion in $\la_0$, albeit in a space with nontrivial measure.

To define Feynman rules, we need the bare vertex. This can be obtained from the Lehmann--Symanzik--Zimmermann (LSZ) formula, as the lowest-order $n$-particle amplitude or, alternatively, as the Fourier transform of the lowest-order $n$-point amputated Green function arising from Eq.\ \Eq{zj}. The answer is 
\ba
V(k_1,\ldots , k_n) &=&\rmi \int\rmd^D x \,\la_{\b-1}(x)\,\rme^{\rmi x\cdot k_{\rm tot}}\label{verte1}\nonumber\\
&=& \rmi\la_0 \left[2\ell c_\b \Gamma(\b)\cos \frac{\pi\b}{2}\right]^D\prod_\mu\frac{1}{|\ell k^\mu_{\rm tot}|^\b}\,,\label{verte2}
\ea
where we used the Fourier transform of $\la(x)$ (see Subsec.\ II.2.3 in Ref.\ \citen{GS}) and $k^\mu_{\rm tot}:=\sum_{i=1}^n k_i^\mu$. For $\b >0$, this equation precisely encodes the notion that momentum is not conserved: the total momentum spreads out of the support $k^\mu_{\rm tot}=0$ as $\prod_\mu |k^\mu_{\rm tot} |^{-\beta}$. In addition, from Eq.\ \Eq{verte2} it follows that the normalization $c_\b = 1/\Gamma(\b)$  does not play any special role, as it is a nonvanishing entire function for $\b>0$. According to \Eq{strips}, the distribution $\prod_\mu |k^\mu_{\rm tot} |^{-\beta}$ is defined over the whole complex plane with the exception of the values $\b = 2 l +1$, $l\in {\mathbb N}$, where it presents  simple poles (one for each direction $k^\mu$). Its Laurent expansion can be easily deduced from Eq.\ \Eq{usefu}, and reads
\be
\prod_\mu\frac{1}{| k^\mu_{\rm tot}|^\b} = \prod_\mu \left[ - \frac{2 \delta^{(2l)}(k^\mu_{\rm tot})}{(2l)! (\b-2l-1)}+\ldots\right],
\ee
where dots denote terms analytic in the $\b\to 2l+1$ limit. However, for such values of $\b$, the poles are canceled by the vanishing prefactors $\cos(\pi\b/2)$ in $V$ (one for each direction), and the neat result is that the vertex \Eq{verte2} is well defined for any $\b>0$. In particular, the limit $\b \to 2 l +1$ gives, with the normalization  $c_\b =1/\Gamma(\b)$,
\be\label{ver}
V(k_1,k_2,k_3)= \rmi\la_0\left[\frac{2\pi (-1)^l }{\ell^{\, 2 l }\, (2l)!}\right]^D\de^{(2 l)}(k_{\rm tot})\, ,
\ee
where $\de^{(2 l)}(k_{\rm tot})= \prod_\mu \de^{(2 l)}(k^\mu_{\rm tot})$. Contrary to Eq.\ \Eq{verte2}, for the special values $\b = 2 l +1$, the support of the vertex is concentrated at $k_{\rm tot}=0$, and  momentum is perturbatively conserved at the quantum level at least if the perturbative series is truncated at any finite order.  This is quite an unexpected feature which, however, has some important caveats with it that we will discuss in the next section. Notice that the result \Eq{ver} can be intuitively understood by taking the naive Fourier transform of a monomial $\la(x)\sim x^n$, which is proportional to $(-\rmi)^n\de^{(n)}(k)$. However, Eq.\ \Eq{ver} has been obtained by an analytic continuation and limiting procedure which automatically select even powers $n$ and, hence, unitary theories.

As a consistency check, if $\b=1$ (so that $l=0$), we get the standard case $V=\rmi\la_0 (2\pi)^D\de(k_{\rm tot})$. The next critical value is $\b=3$ (i.e., $l =1$), for which
\be\label{vert}
V=\rmi\la_0\left(-\frac{\pi}{\ell^2}\right)^D\de''(k_{\rm tot})\,,
\ee
where we have the second momentum derivative of the Dirac distribution, $\de''(k_{\rm tot})=\prod_\mu\de''(k^\mu_{\rm tot})$. Momentum does not spread out for $\b = 2 l +1$, $l\in {\mathbb N}$.


\section{Examples of Feynman Diagrams}

We shall consider, as an example, the case of cubic potential, $n=3$. For general $\b$, it is very difficult to set up a viable perturbation theory.  The main problem is that the  ``volcano-like'' vertices \Eq{verte2} do not compose with a group law under convolution: composing free propagators and vertices behaving like $|k|^{-\beta}$ does not reproduce amplitudes behaving like $|k_{\rm tot}|^{-\b}$ in the total momentum. Therefore, even if possible in principle, perturbation theory is hardly manageable. The case $\b=2l+1$ is considerably easier.  At least, composing vertices with conserved momenta gives amplitudes where momentum is conserved at any given order. Tadpole diagrams, which could provide a nonvanishing vacuum expectation value for a single field, can be systematically removed by introducing a counterterm $\lambda (x) Y \varphi (x)$ in the Lagrangian, just like in a standard $\varphi^3$ theory \cite{Sre07}.\footnote{This happens because tadpoles descend from a propagator closing around the same vertex, and in our case the free propagator is identical to the standard one.}

Still, one may need further simplifications after a certain point. For instance, consider the $2\rightarrow 2$ particle amplitude in the cubic theory with $\b=3$ ($l=1$). At the lowest order, the amplitude is given by the sum of the three one-particle-exchange diagrams
\ba
\vphantom{1}&&\parbox{10.5cm}{\includegraphics[width=10cm]{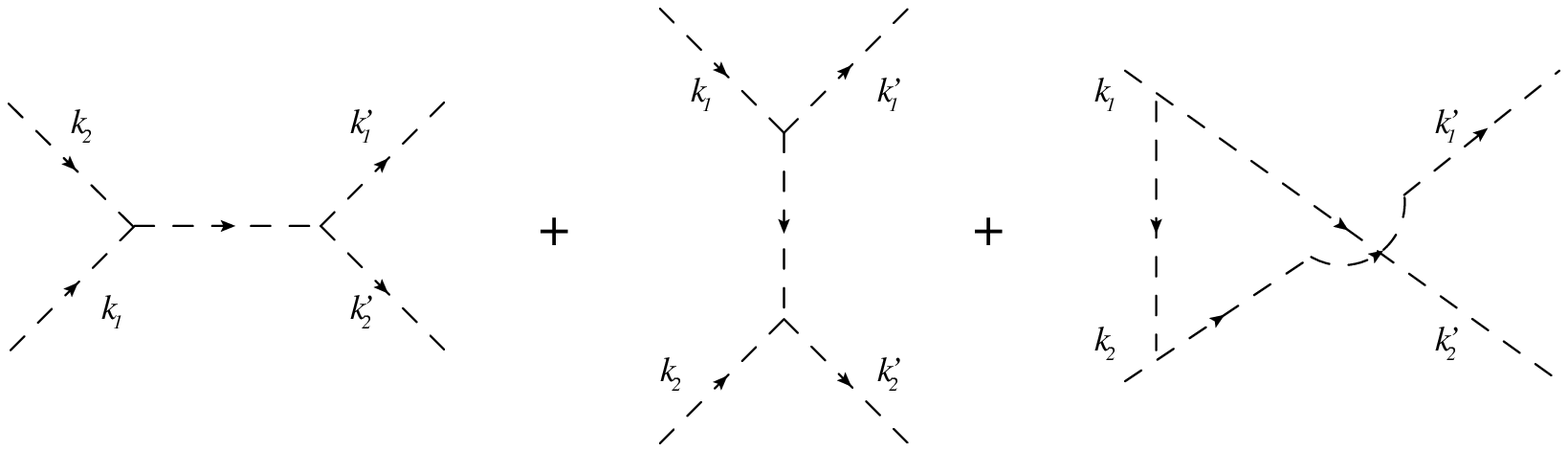}} = \langle{\rm f}|{\rm i}\rangle\nonumber\\
&&\qquad = \rmi \la_0^2\left(\frac{\pi}{2 \ell^4}\right)^D \int\frac{\rmd^Dk}{k^2+m^2-\rmi\e}\left[\de''(k+k_1+k_2)\de''(k+k_1'+k_2')\right.\nonumber\\
&&\qquad\qquad\left.+\de''(k+k_1-k_1')\de''(k+k_2'-k_2)+\de''(k+k_1-k_2')\de''(k+k_1'-k_2)\right].\nonumber
\ea
At this point, however, it is not possible to reorganize the terms into a single double-derivative delta $\de''(k_1+k_2-k_1'-k_2')$, as the form of the vertex would suggest.\footnote{Here and in the following, there should be no confusion between primes denoting momentum derivatives and those indicating outgoing momenta.} This is due to the presence of the derivatives which, upon integrating by parts, are unloaded onto various mixed contributions where momenta along different directions are entangled. A simplification occurs either in the least interesting case $D=1$ or when the coupling $\la$ depends only on one direction, for instance $\mu=0$. For this configuration (in the language of multiscale geometry, corresponding to an anisotropic measure where only one of the directions is fractional), $\la(x^0)=\la_0 (x^0/\ell)^2/2$, the vertex in \Eq{vert} should be replaced by
\be
V= -\rmi \frac{\lambda_0}{2 \ell^2} (2\pi)^D \Delta(k) =   -\rmi\frac{\lambda_0}{2 \ell^2} (2\pi)^D \delta''(k^0)\delta ({\bf k})\,,
\ee
and one can show that
\ba
\langle{\rm f}|{\rm i}\rangle &=&-\rmi\frac{\la_0^2(2\pi)^D}{4\ell^4}\nonumber\\
&&\times\frac{\rmd^2}{\rmd k_1^{0\,2}}\left[ \Delta(k_1+k_2-k_1'-k_2')\left(\frac{1}{s-m^2}+\frac{1}{t-m^2}+\frac{1}{u-m^2}\right)\right]\!,
\ea
where $s=-(k_1+k_2)^2=-(k_1'+k_2')^2$, $t=-(k_1-k_1')^2=-(k_2-k_2')^2$ and $u=-(k_1-k_2')^2=-(k_2-k_1')^2$ are Mandelstam variables. The corresponding amplitude for $\b=1$ would be without the second derivative in front and with $\Delta$ replaced by $\delta$.

The self-energy diagram  presents similar properties. Still limiting our attention to the anisotropic case above, one can show that\footnote{Standard notation for the self-energy function (here not viable due to the presence of derivatives) is recovered upon $k'$ integration.} 
\ba
\parbox{4cm}{\includegraphics[width=3.5cm]{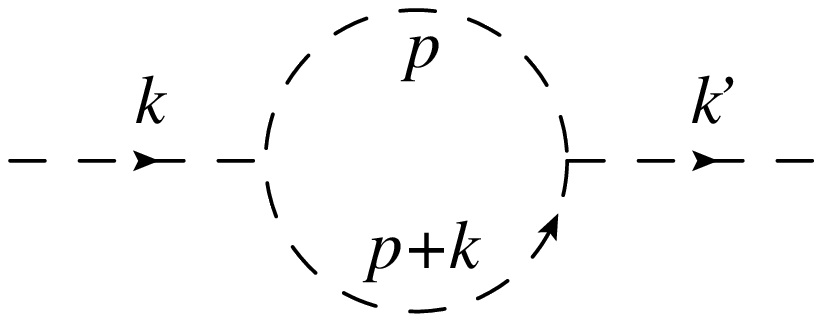}}  &=& \rmi \Pi(k,k')\nonumber\\
&=&\rmi(2\pi)^D\de({\bf k-k'}) \frac{\rmd^2}{\rmd k^{0\,2}}\left[\de''(k^0-{k^0}')\tilde\Pi(k^2)\right],\label{buda}
\ea
where ${\bf k}$ denotes spatial components of the momentum and
\be\label{tip}
\tilde\Pi(k^2):=\frac{\la_0^2}{8\rmi\ell^4}\int\frac{\rmd^Dp}{(2\pi)^D}\frac{1}{(p^2+m^2-\rmi\e)[(k+p)^2+m^2-\rmi\e]}\,
\ee
is (up to a constant factor $1/(4\ell^4)$) the same self-energy function as in the standard theory with constant coupling. The integral \Eq{tip} can be performed explicitly (see, for instance, Ref.\ \citen{Sre07}).

The bubble diagram \Eq{buda} is the simplest graph  to inspect the renormalization properties of the theory. Consider the block made by the self-energy diagram multiplied by a free propagator in $k'$ and integrated over the internal momentum $k'$:
\ba
\parbox{3cm}{\includegraphics[width=2.5cm]{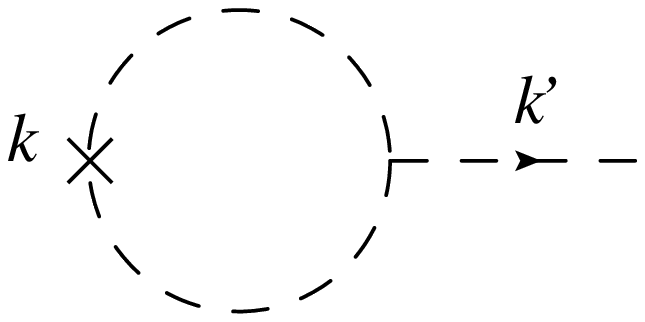}}  &=& \int\frac{\rmd k'}{(2\pi)^D}\rmi\Pi(k,k')\frac{-\rmi}{{k'}^2+m^2-\rmi\e}\nonumber\\
&=&\frac{\rmd^2}{\rmd k^{0\,2}}\left[\tilde\Pi(k^2)\frac{\rmd^2}{\rmd k^{0\,2}}\frac{1}{k^2+m^2-\rmi\e}\right]=: \p^2(\tilde\Pi\, \p^2 \tilde G_0)\,,\label{bloc}
\ea
where in the last member we introduced a self-explanatory minimalistic notation. This block can be iteratively composed to give the Dyson series for the full quantum propagator $\tilde G$, that can be formally summed:
\be
\tilde G = \tilde G_0+A \tilde G_0+A(A \tilde G_0)+\dots=[1-A]^{-1}\tilde G_0\,,\qquad \label{ga}
\ee
where $A:=\tilde G_0\p^2(\tilde\Pi\p^2\,\cdot\,)$.  
Although the propagator is expressed in a closed form, it is not so manageable due to the fact that $A$ is a mixed (multiplicative and differential) operator, while in the standard case $A$ would be just multiplicative. Perhaps, more insight could be gained by finding either some different integral representation of Eq.\ \Eq{ga} or a functional basis where the action of the $A$ operator is simple, and expanding eventually $\tilde G_0$ around such a basis.

The support of the general finite-order perturbative Feynman diagram is at $k_{\rm tot}=0$, just like the bare vertices in Eqs.\ \Eq{ver} and \Eq{vert}, and the total momentum is conserved. In general, it is not guaranteed that an infinite sum of derivatives of increasing order of the Dirac distribution has the same support of the original delta. More concretely, write this sum as the functional $g(x)=f(\p)\,\de(x)=\sum_n c_n \de^{(n)}(x)$, for some coefficients $c_n$. Applying it to some test function $\Phi(x)$, the result is $(g,\Phi)=[f(-\p)\Phi](0)$. If, for instance, $f(\p)=\exp(-a\p)$, then $(g,\Phi)=\Phi(a)$ and $g(x)=\de(x-a)$; namely, an infinite sum of distributions with support at $x=0$ gives, as a result,  a distribution with support at $x=a$. Such a mechanism could manifest only if an infinite sum of derivatives of increasing order is involved. If something similar happened in the present field model, then perturbation theory would be unpredictive because, at any given finite order, one would be throwing away terms which are not subleading with respect to the others. In this worst-case scenario, one could hope to quantize and renormalize the theory only with nonperturbative techniques. Clearly, the source of the trouble is the appearance of higher and higher derivatives in higher-order vertex expansions, due to nonconservation of the form of the vertices order by order: the theory is nonlocal \emph{de facto}. Here we will not attempt to check whether such a perturbative disaster occurs also in the type of sums appearing in our model. This might not necessarily be the case, in fact. The weak point of the above naive nonlocal example is that it does not take into account the actual structure of the field theory, where the individual terms of infinite sums such as \Eq{ga} are integrals such as \Eq{bloc}. In general, the sum and the integration operations do \emph{not} commute, which however makes the problem difficult to assess.

With this in mind, we can examine the degree of divergence of the diagram \Eq{buda} and compare it with the one in standard field theory. The $D=6$ example has been worked out in, e.g.\ Chap.\ 14 of Ref.\ \citen{Sre07}. Via dimensional regularization, one can show that, up to some multiplicative constant, $\tilde\Pi(k^2) \sim (k^2+m^2) \ln (k^2/m^2)$ for large $|k^2|$. In our case, we have an extra second-order derivative in $k^0$: this produces three terms from Eq.\ \Eq{buda}, proportional to $\de''''\tilde \Pi$, $\de'''\tilde \Pi'$ and $\de''\tilde \Pi''$. Of these, the first dominates over the other two and has the same degree of divergence as the usual theory. Barring the unforeseen occurrence of cancellations order by order, the degree of divergence of individual graphs seems to be the same of the standard theory.

Equation \Eq{ga} also gives the effective dispersion relation for large momenta. In the coupling expansion up to $O(\la_0^2)$, we keep only the first two terms of the Dyson series. In the large-$k$ limit, this yields
\be
\tilde G \approx \tilde G_0+A \tilde G_0\sim \frac{1}{k^2}+C\left[\frac{\ln k^2}{k^6}+O(k^{-6})\right]\approx \frac{1}{k^2-C\ln k^2/k^2}\,,
\ee
where $C=O(\la_0^2/\ell^4)$ is a constant and we used a heuristic notation to represent the modified momentum dependence of the propagator in a given direction. Thus, in the ultraviolet ($k\to\infty$) the correction term is subdominant with respect to the usual one (even more so in $D=4$), and we avoid the fine-tuning problem mentioned in the introduction, which can be seen as an observationally unacceptable enhancement of Lorentz-symmetry violation \cite{CPSUV,CPS}. This happens, in general, in models where the dispersion relation acquires terms which dominate at small scales, as for instance in Lifshitz-type field theories \cite{IRS}. However, there are other quantum-gravity models which can bypass that argument \cite{GRP}.


\section{Discussion}

A spacetime-dependent nondilatonic coupling manifestly violates time and space translation invariance. As a consequence, momentum is not conserved in a generic field theory with nonconstant nondynamical couplings. We have shown that, for a special class of models, it may be possible to recover momentum conservation at the quantum level at least at finite order in perturbation theory. Spacetime-dependent couplings have been recently introduced in the context of fractional field theories \cite{AIP,frc6,fra4,frc1,frc2,fra6}. The ultimate interest in the latter is the possibility to tune, within the same geometry, different perceived dimensions and renormalization properties at different scales. Such a scale dependence is manifest in effective regimes of quantum geometries such as in asimptotic safety, causal dynamical triangulations and noncommutative spacetimes. Multiscale (in particular, multifractional) field theories can provide an effective description of these regimes \cite{ACOS,fra7}.

The results of this paper indicate that, although the details of the Feynman diagrams of the theory with varying coupling can considerably differ from the usual scalar theory, their degree of divergence is the same. This immediately affects the previously mentioned fractional field theories with weighted Laplacian. Their renormalization properties, it seems, is \emph{not} improved with respect to the standard theory. One might object that the evidence collected in favour of this conclusion is circumstantial. We did not extract concrete information on the degree of divergence of Feynman diagrams except in the case $\b=2l+1$ which, applied to Eq.\ \Eq{ab}, corresponds to a fractional charge
\be\label{aln}
\a=1-\frac{4l}{n-2}\,.
\ee
The actual case of study was limited to $l=1$ and $n=3$, for which $\a=-3$ seems not to correspond to a conventional fractional geometry (where $\a\geq 0$). However, this configuration is actually an anisotropic one with fractional time direction, having a Hausdorff dimension $\dh=\a+D-1=D-4$, which is exactly zero in four dimensions. This value, albeit extreme, still belongs to a sensible nontrivial geometry. Moreover, values within the range $0\leq \a\leq 1$ are allowed for suitable (low) $l$ and (high) $n$; for instance, $l=1$ corresponds to $\a=0,1/5,1/3,3/7,1/2,\dots$ for $n=6,7,8,9,10,\dots$\,. 

Another argument, suggesting that the renormalizability of this class of fractional models is basically the same as that of the standard theory, is a revision of the power-counting argument of Ref.\ \citen{frc2}. The main agent lowering the superficial degree of divergence of Feynman diagrams was expected to be the momentum integration in loops. The measure there is $\rmd^Dk\,w(k)$, where the weight $w(k)$ is such that the scaling dimension of the measure is smaller than $D$. However, when coupled with the full expression with two fractional phases $\bE(k,x)=\rme^{\rmi k\cdot x}/\sqrt{w(k) v(x)}$ (such as in propagators), the latter include two factors $w^{-1/2}$, which cancel the weight in the measure. Thus, the degree of divergence of momentum integrals remains the same as in the integer field theory. In this paper, we have found the actual degree of divergence of some diagrams, which differs with respect to the above power-counting argument but essentially agrees with its main conclusion. Yet another, more intuitive way to understand this point is to notice that the free multiscale propagator in position space is of the form $G_{v,{\rm free}}(x,y)=G_0(x-y)/\sqrt{v(x)v(y)}$ for any factorizable positive semidefinite measure $v$ \cite{frc6}. Therefore, the divergence of $G_{v,{\rm free}}(x,y)$ at coincident points $x\sim y$ is solely determined by the usual propagator $G_0(x-y)$ and not by the prefactor $\sim 1/v(y)$.\footnote{The only delicate points where this argument may fail are those corresponding to the measure singularities ($y=0$ in the fractional case), where the above expression for $G_{v,{\rm free}}(x,y)$ is ill defined. However, we do not expect them to modify the main conclusion.}

This does not exclude that other classes of multifractional spacetimes, endowed with a different set of symmetries, have improved renormalization properties. However, also the class with $q$-Laplacians \cite{frc6} is likely not to work in this respect. The reason is the same as before. In this class, the theory in position space is the same as the standard one upon replacing coordinates $x^\mu$ with the composite multiscale object $q^\mu(x^\mu)=\int\rmd x^\mu\,v(x^\mu)$. The nontriviality of the theory is then guaranteed by the choice of physical momenta, which are conjugate to $x$ and not to $q$ (the latter has anomalous scaling, $q(\la x)\neq \la q(x)$). The free propagator is $G_{q,{\rm free}}(x,y)=G_0[q(x)-q(y)]$ and its behaviour at $x\sim y$ is the same as the standard theory. For instance, in the massless case $G_0[q(x)-q(y)]\propto |q(x)-q(y)|^{2-D}\sim |v(y)(x-y)|^{2-D}$ upon Taylor expanding around $x=y$, and at coincident points inverse powers of $q(x)-q(y)$ will diverge as inverse powers of $x-y$. Although we have not discussed this quantum theory in momentum space, its basic renormalization properties can be inferred from position space, and the above argument may be regarded as robust.

In both classes of multiscale theories (with weighted or $q$ Laplacians), the spectral dimension of spacetime is suppressed at small scales \cite{frc7}. In general, especially in quantum gravity, it is believed that a lower spectral dimension in the ultraviolet is intimately related to improved renormalization properties. Here, however, we have provided one fully worked-out counter-example (and sketched another one) where a renormalization analysis shows how this expectation is \emph{not} fulfilled. Dimensional flow in quantum gravity does not guarantee ultraviolet finiteness. On a positive side, if a standard field theory is renormalizable, the same argument tells us that its modification to a multiscale geometry will share the same property. Of course, all of this is subject to verification of the caveats on the perturbative expansion discussed in the text.

The interest in fractional theories is not jeopardized anyway, for two reasons. First, varying-coupling scenarios find a novel realization and interpretation in the context of multiscale complex systems, which may stimulate thinking into fresh directions with respect to traditional dilaton-like and varying-speed-of-light models \cite{frc8}. Second, the tuning between the fractional charge $\a$ and the degree $n$ of the potential (Eq.\ \Eq{aln}) recovers momentum conservation at the quantum level. At the quantum level, the Noether current is an operator composed of particle fields at the same point, which, due to ultraviolet divergences, needs both a regularization and renormalization procedure to be defined. After the removal of the regulators, these procedures may break the validity of the classical equations leading to derive the current. This is the usual case of anomalies (see, e.g., Ref.\ \citen{Col84}). In our framework, on the other hand, even before regularizing and renormalizing, an extra symmetry, classically broken, is restored for special behaviours of the coupling distribution $\la(x)$, tuned with respect to the interaction, at least at a given finite perturbative order. This unusual phenomenon and its physical interpretation should deserve further attention.


\section*{Acknowledgments}
The work of G.\ Calcagni is under a Ram\'on y Cajal contract.


\end{document}